**Unpacking the Essential Tension of Knowledge Recombination: Analyzing the Impact of Knowledge Spanning on Citation Counts and Disruptive Innovation**

Cheng-Jun Wang [a], Lihan Yan [b], Haochuan Cui [c*]

[a] Collaboratory of Computational Communication,

School of Journalism and Communication, Nanjing University, Nanjing, China.

Address: 163 Xianlin Road, Qixia District, Nanjing, Jiangsu Province, 210023, P.R.China.

Email: wangchengjun@nju.edu.cn

[b] Collaboratory of Computational Communication,

School of Journalism and Communication, Nanjing University, Nanjing, China.

Address: 163 Xianlin Road, Qixia District, Nanjing, Jiangsu Province, 210023, P.R.China.

Email: yanlihan9829@gmail.com

[c] School of Systems Science, Beijing Normal University, Beijing, China.

Address: 19 Xinjiekouwai Street, Haidian District, Beijing, 100875, P.R.China.

Email: haochuancui@mail.bnu.edu.cn

[*] Corresponding author. E-mail addresses: haochuancui@mail.bnu.edu.cn (H. Cui)

# Unpacking the Essential Tension of Knowledge Recombination: Analyzing the Impact of Knowledge Spanning on Citation Counts and Disruptive Innovation

Cheng-Jun Wang, Lihan Yan, Haochuan Cui

## Abstract

Drawing on the theories of knowledge recombination, we aim to unpack the essential tension between tradition and innovation in scientific research. Using the American Physical Society data and computational methods, we analyze the impact of knowledge spanning on both citation counts and disruptive innovation. The findings show that knowledge spanning has a U-shaped impact on disruptive innovation. In contrast, there is an inverted U-shaped relationship between knowledge spanning and citation counts, and the inverted U-shaped effect is moderated by team size. This study contributes to the theories of knowledge recombination by suggesting that both intellectual conformism and knowledge recombination can lead to disruptive innovation. That is, when evaluating the quality of scientific research with disruptive innovation, the essential tension seems to disappear.

## Introduction

Innovation paves the way for scientific research (Koestler, 1964; Nelson & Winter, 1982; Schumpeter, 1934). However, according to the seminal work of Thomas Kuhn (1977), there is an essential tension between tradition and innovation. Kuhn (1977) asserts that productive tradition and risky innovation are indispensable for scientific research. In line with this logic, prior studies keep finding that knowledge spanning has an inverted U-shaped impact on the impact or success of scientific research (Foster et al., 2015; Guan et al., 2017; Liu et al., 2020), patents (Zhang et al., 2019), musicals (Uzzi & Spiro, 2005), songs (Askin & Mauskapf, 2017), and questions (Shi et al., 2021). These empirical findings of the essential tension illustrate the innovators' dilemma in scientific research. However, prior research primarily focuses on the citation counts of scientific research rather than disruptive innovation (Guan et al., 2017; S. Wang et al., 2021; Zhu et al., 2021). Although citation counts are simple and intuitive, they suffer from a series of unintended problems, such as ignoring negative citations (Catalini et al., 2015), increasing citation inequality (Nielsen & Andersen, 2021), and encouraging the strategic choice of conservative research (Foster et al., 2015). Therefore, we formulate the key puzzlement of this research as follows: if we evaluate the quality of scientific research with disruptiveness, does the essential tension (i.e., the inverted U-shaped impact of knowledge spanning) still exist?

If the logic of scientific research driven by citation counts is not balanced by fostering disruptive scholarship, the progress of scientific research would be trapped in the existing canon. Foster and his colleagues (2015) find that high-risk innovation strategies are rare in biomedical research. Rzhetsky et al. (2015) analyze biomedical research over 30 years, and further find that biomedical researchers pursue conservative research strategies, which slows scientific advances. By analyzing the Web of Science articles published between 1954 and 2014, Wu et al. (2019) show that the disruptiveness of scientific research decreases over time.

Similarly, Chu and Evans (2021) find that scientists in a field are more inclined to develop existing ideas than propose disruptive (or highly disruptive) ones. Although the number of published scientific research increases over time, the central ideas in a field have not been changed; In contrast, canonical research has a durable dominance in terms of garnering new citations (Chu & Evans, 2021). In all, using citation counts to evaluate the quality of scientific research tends to encourage the strategic selection of traditional research.

To address this theoretical puzzlement, we draw our research on the theories of knowledge recombination (Nelson & Winter, 1982) and category spanning (Hannan, 2010; Hannan & Freeman, 1977a; Hsu et al., 2009). We define *knowledge spanning* as one form of knowledge recombination in terms of knowledge space. Using the computational methods of word embeddings and network analysis, we model the knowledge space as a high-dimensional geometric space or a tree. Meanwhile, we trace how each scientific research spans the boundary of knowledge space and quantify its extent of knowledge spanning. Specifically, we calculate two types of measures of knowledge spanning: geometric distance and network distance. According to the scope and direction of recombination, we further classify them into internal knowledge spanning and external knowledge spanning. *Internal knowledge spanning* refers to the scope covered by the categories of scientific research in the knowledge space. The larger the coverage in the knowledge space, the greater the internal knowledge spanning. Thus, the internal knowledge spanning measures the step size of scientific research. However, it fails to capture the direction of research that is more critical for scientific progress. The research aligning with the direction of the paradigm tends to be conventional. While the research deviates from the direction of the paradigm tends to be more disruptive. Following this logic, we define *external knowledge spanning* as the extent to which scientific research deviates from the mainstream paradigm. Finally, we analyze the impact of knowledge spanning on citation counts and disruptive innovation.

We claim that the existence of the essential tension between tradition and innovation depends on how we evaluate the quality of scientific research. Testing the hypotheses using the APS data set (American Physical Society, 2021), this present research contributes to prior research in the following aspects: First, we reformulate the puzzlement of the essential tension by distinguishing scientific credit into citation counts and disruptive innovation. Surprisingly, our findings show that although knowledge spanning has an inverted U-shaped impact on the popularity of scientific research, it has a U-shaped impact on the disruption of scientific research. Thus, the essential tension disappears when we evaluate the quality of scientific research with disruptive innovation. Both the lower level and higher level of knowledge spanning can breed more disruptive innovation. Therefore, disruptive innovation prefers both lower and higher level of knowledge spanning than a mixture of them. Second, given that team size plays a crucial role in scientific research, we have also considered the impact of team size. On the one hand, we support prior research on the main effect of team size (Wu et al., 2019; Wuchty et al., 2007). On the other hand, we further show that team size moderates the impact of knowledge spanning on citation counts. In all, this study contributes to the theories of knowledge recombination by unpacking the essential tension proposed by Kuhn (1977).

The other sections of this study are organized as follows: First, we review the literature and establish our theoretical framework based on knowledge recombination and team science. Second, based on this theoretical framework, we propose research hypotheses to answer the research puzzlement. Third, we introduce the dataset, measures, and models and test our hypotheses. Finally, we discuss our findings, limitations, implications, and conclusions.

## The Essential Tension and Knowledge Spanning

We construct our theoretical framework based on the idea of the essential tension proposed by Thomas Kuhn (1977). In the seminal work titled *The Essential Tension*, Kuhn (1977) argues that both convergent and divergent thinking are essential to the progress of scientific research. Thus, there is a tradeoff between productive tradition and risky innovation. On the one hand, scientists are trained to be puzzle-solvers who follow tradition and selectively ignore most unexpected discoveries; On the other hand, scientists are aware that their long-term reputation comes from their innovative discoveries (Kuhn, 1977). The lack of either tradition or innovation is harmful to the development of science.

The legitimacy of normal science comes from the exemplars established by prior scientific revolutions (Kuhn, 1962). Kuhn (1962) conceptualizes these exemplars as the paradigm. The paradigm is like a compass that guides the scientific community on a planned route. We regard scientific tradition as a big ship carrying sailors who share the same vision about the direction of their expedition. According to Thomas Kuhn (1962), normal science aims to extend the scope and precision of scientific knowledge rather than to establish novel facts or theories. The researchers of normal science are puzzle-solvers who limit their scopes to "*achieve the anticipated in a new way*" (Kuhn, 1996, p. 36). The tasks of normal science are restricted to three categories: precisely determining the facts, matching facts with theories, and articulating theories (Kuhn, 1996, p. 34). In addition, Kuhn (1962) proposes that the development of normal science is bound to undergo a process of professionalization. Professionalization is manifested in the establishment of sophisticated equipment, the improvement of skills, and the refinement of concepts and vocabulary (Kuhn, 1962). In all, tradition fixes the route of making scientific discoveries. As a result, it is reasonable to contend that there is less possibility of making discoveries on the road of such traditional research.

However, and surprisingly, although Kuhn denies the role of innovation in normal science, he admits that innovation is an important feature of scientific research. "*New and unsuspected phenomena are, however, repeatedly uncovered by scientific research, and radical new theories have again and again been invented by scientists*" (Kuhn, 1996, p. 52).

Kuhn (1962) further argues that the research under a paradigm can effectively induce new paradigm shifts. Discoveries emerge from old theories (Kuhn, 1962). In other words, only the innovations rooted in the scientific tradition can break the tradition and give rise to scientific revolutions. He attributes this unintended outcome to the professionalization of normal science that provides detailed information, improves the accuracy of observations, and makes the existence of anomalies more conclusive (Kuhn, 1996, p. 65). As a result, paradigm change happens, new phenomena are discovered, and novel theories are constructed. According to this logic, paradigm change is only possible when normal science has experienced or even finished the professionalization process.

We theorize the essential tension from the perspective of knowledge recombination (Koestler, 1964; Nelson & Winter, 1982; Schumpeter, 1934). Our understanding of innovation comes from Joseph Schumpeter's exploration of the impact of technological innovation on economic development. Schumpeter considers innovation as implementing "new combinations" (Schumpeter, 1934, pp. 65–66). In light of Joseph Schumpeter's idea, Nelson and Winter further argue that the creation of novelty in science originates from the "recombination of conceptual and physical materials" (Nelson & Winter, 1982, p. 130). Similarly, Koestler (1964) coins the term bisociation to describe this innovation process and asserts that creative ideas lie at the intersection of two frames of thought. Further, when the degree of knowledge recombination is relatively small, scientific research is more traditional; On the contrary, scientific research is more innovative. Thus, the degree of knowledge recombination depicts the tension between tradition and innovation.

Bourdieu (1975) views scientists as a function of their positions within the scientific field. The scientific field is one form of social field featured by the competitive struggle for scientific stakes or authority. Scientists make strategic choices to accumulate scientific capital (e.g., peer recognition) and maximize productivity. Since innovation is risky and would be punished for deviating from the dominating paradigm, a disposition toward tradition rather than innovation is a more rational choice. In this sense, innovation is more like a gamble. Foster et al. (2015) further illustrate the mechanisms of the essential tension based on Bourdieu's field theory of science. By analyzing how biomedical researchers establish links between chemicals, they classify scientific strategies into three categories: consolidation, bridge, and jump. Although innovative research is more likely to achieve a higher impact (measured by citation counts) than a conservative one, it is rarely adopted by scientists because the reward of risky innovation can not offset the loss of rejection (Foster et al., 2015).

In light of the studies of category spanning (Hannan, 2010; Hannan et al., 2007; Hannan & Freeman, 1977b; Hsu et al., 2009), we further operationalize knowledge recombination as knowledge spanning across categories. Categorization plays an essential role in scientific research. Scientific knowledge is categorized into different domains, and scientists are categorized into different communities. For example, keywords are widely used to organize, locate, and discover scientific research in the knowledge space. The keywords of scientific research constitute its coordinates in the knowledge space. Using keywords, readers can easily locate research in the knowledge space. Moreover, keywords can also indicate the scope of knowledge space a study covers and to what extent this study deviates from mainstream research. The categories of scientific research act as boundaries that hinder recombination in the knowledge space. Thus, knowledge spanning can be viewed as one kind of category spanning, reflecting the distance of boundary spanning in the knowledge space. According to prior research, our definition of knowledge spanning quantifies the novelty of scientific research (J. Wang et al., 2017).

**Research Hypotheses**

According to prior research (Askin & Mauskapf, 2017; Foster et al., 2015; Guan et al., 2017; Shi et al., 2021; Uzzi & Spiro, 2005; Zhang et al., 2019), the essential tension proposed by Kuhn (1977) suggests that there is a critical point between tradition and innovation. Specifically, there is an inverted U-shaped relationship between knowledge spanning and its impact. The inverted U-shape exists because there is negative feedback with the increase of knowledge spanning. On the one hand, if a study only spans a short distance in the knowledge space, it would be punished because it is boring, and the readers are less likely to read it. On the other hand, if a study spans a long distance in the knowledge space, it would be penalized because it is difficult to interpret, and the readers are unwilling to read it. Scientists need to be close to the critical point if they want to be successful. In contrast, their achievements will be constrained if they deviate from the critical point. Although there is an inevitable contradiction between tradition and innovation, successful scientists can maintain the essential tension between tradition and innovation. Thus, the essential tension provides a cost-benefit framework for understanding the impact of scientific innovations. For example, Foster et al. (2015) find an inverted U-shaped relation between the proportion of scientific strategy (i.e., jump, new bridge, repeat bridge, new consolidation, repeat consolidation) per article and citation counts. Yan et al. (2020) find that both a paper's new combinations (i.e., new pairs of knowledge elements in a related research area) and new components (i.e., new knowledge elements that have never appeared in a related research area previously) have an inverted U-shaped effect on its citation count. We posit that the existence of the essential tension between tradition and innovation depends on how we evaluate the quality of scientific research. The essential tension holds when we evaluate scientific research merely by citation counts. Thus, we propose the first hypothesis as follows:

*H1*: knowledge spanning has an inverted U-shaped impact on citation counts.

The theory of category spanning explains the negative feedback of knowledge spanning from two aspects (Keuschnigg & Wimmer, 2017). First, knowledge spanning reduces researchers' niche fitness. According to the degree of knowledge spanning, researchers can be distinguished into specialists and generalists. Compared with specialists, the generalists or category spanners tend to be distracted and lose focus. Thus, it is more difficult for the generalists to accumulate experience and foster niche fitness; Second, knowledge spanning increases the audience's confusion about scientific research. Scientists have a stable expectation of scientific research. From the perspective of symbolic interaction theory, scientists tend to understand scientific research from the perspective of their mental images (Becker, 1998). In addition to common sense, the education of scientists plays a vital role in constructing our mental images (Kuhn, 1977). Both common sense and domain knowledge reflect the logic of the dominant paradigm. Nevertheless, knowledge spanning means deviating from the paradigm and our mental images. Consequently, knowledge spanning increases cognitive difficulties in understanding scientific research.

Nonetheless, by transforming the way of evaluation, we argue that it is possible to eliminate the essential tension (i.e., the negative feedback of knowledge spanning) and transfer scientific research into a new track of innovation. In this study, we consider an alternative evaluation of citation counts—the disruptiveness of scientific research (Funk & Owen-Smith, 2017; Wu et al., 2019). The intuition of disruptiveness is that disruptive innovation shifts scientists' attention away from prior research (Funk & Owen-Smith, 2017). Using Plato's analogy, the mainstream paradigm is like the sun, emitting dazzling light. It is usually difficult for people to escape the rational sunshine. However, disruptive research can block the light of the sun, just like what the moon does during a solar eclipse. Disruptive innovation means a break or discontinuity with the current paradigm (Lin et al., 2022). In short, disruptive innovation is a departure from the tradition, which blocks the influence of the dominant paradigm and forms a break or shift of collective attention.

Knowledge spanning can help scientific research leap in the knowledge space. On the one hand, knowledge spanning can improve the generalists' niche fitness in an ecosystem that encourages innovations. On the other hand, disruptiveness means disengaging from the mainstream paradigm. Using disruptiveness as an evaluation metric helps build a new mental image that prefers novelty. Therefore, when we evaluate scientific research by its disruptiveness, the punishment for knowledge spanning could be reduced and even eliminated. We posit that the negative feedback of knowledge spanning will disappear. Based on the logic above, we propose the following research question:

*Q1*: what is the relationship between knowledge spanning and disruptive innovation?

Larger teams can receive more citations than smaller teams because the professionalization logic of normal science is more pronounced for larger teams (Kuhn, 1962). First, large teams are more likely to converge to incumbent paradigms (Wu et al., 2019). Scientific teams are assembled to breed and test new ideas, but both team size and the impact of scientific teams have been increasing over time (Wuchty et al., 2007). When teams get larger and the individual opinions get more diverse, reaching a consensus on the choice of research questions becomes more difficult to. Thus, larger teams tend to be more traditional than disruptive. Although the mission of innovators is to produce diverse ideas, deviating from the mainstream paradigm will be punished. Since disruptive research is often immature, the risks faced by smaller teams are even more deadly. Further, larger teams usually focus on the hot issues related to the dominating paradigm. Therefore, they can reap more attention and have a larger capability to attract citations than smaller teams. Second, through the division of labor, the members of larger teams can give free rein to their expertise, develop their strengths, and avoid their weaknesses. Thus, the research efficiency of larger teams is higher than smaller teams (Zhu et al., 2021). Third, larger teams are often assembled around an experienced and prestigious researcher who occupies a higher position in the scientific field. Consequently, larger teams are generally more influential than smaller teams. In addition, larger teams can mobilize more resources (Zhu et al., 2021). Together, larger teams have a stronger capability to deal with risks and get more citations than smaller teams.

Compared with larger teams, smaller teams are more likely to disrupt science and technology (Wu et al., 2019). Smaller teams are like smaller boats that can flexibly turn around to explore diverse directions. If the comparative advantage of larger teams is to consolidate the incumbent paradigm through division of labor, the comparative advantage of smaller teams is to disrupt science through knowledge spanning. Smaller teams tend to disrupt science by searching deeper into the past, and larger teams tend to develop science by working on popular issues (Wu et al., 2019). The average age of the references cited by smaller teams is generally larger compared with larger teams. In contrast, the popularity of the references cited by smaller teams is weaker compared with larger teams (Wu et al., 2019). Based on the logic above, we propose the hypotheses as follows:

*H2*: team size has a positive effect on citation counts (*H2a*) and a negative impact on disruptiveness (*H2b*).

The influence of knowledge spanning depends on how scientists collaborate. Team size plays a crucial role in the social interaction process among team members. For example, the structure of the scientific team is determined by team size. Using the individual contribution data, Haeussler and Sauermann (2020) find that larger teams have a larger share of specialists and a smaller share of generalists than smaller teams. Compared with the share of authors participating in empirical analysis, the share of authors participating in conceptualization and writing decreases significantly with the increased team size.

As we have argued above, larger teams are characterized by a more potent logic of professionalization than smaller teams. On the one hand, the professionalization of larger teams would limit scientists' horizons and hinder disruptive research. For example, Park and his colleagues (2021) quantify the narrowing scope of scientists by the decline in the cited references' diversity, the increase in self-citation, and the increase in cited references' mean age. They argue that the narrowing scope of scientists places restrictions on the disruptiveness of scientific research, which is why the disruptiveness of scientific research has been declining in the past six decades (Park et al., 2021). On the other hand, the professionalization of larger teams can make them more productive and influential. Arguably, team size can amplify the impact of knowledge spanning on citation counts. Focusing on the scientific impact of the scientist, Zhu, Liu, and Yang (2021) find that team size has an inverted-U shaped relationship with research impact (i.e., weighted average citations) and research variety (i.e., the number of disciplinary categories covered by a focal scholar) can moderate this nonlinear relationship. In comparison, we focus on the scientific impact of the individual scientific research, the predictor is the extent of knowledge spanning of the scientific research, and the moderation variable is the team size. Thus, we propose the following hypotheses:

*H3*: team size moderates the impact of knowledge spanning on citation counts (*H3a*) and disruptiveness (*H3b*).

## Methods

**Data**

We use the American Physical Society (APS) dataset to quantify knowledge spanning as well as team size and examine their impact on scientific research. Founded in 1899, APS publishes over ten journals of physical review series, such as Physical Review A (PRA), Physical Review B (PRB), Physical Review C (PRC), Physical Review D (PRD), Physical Review E (PRE), Physical Review X (PRX), Physical Review Letters (PRL), Physical Review Special Topics-Physics Education Research (PRSTPER), Physical Review Special Topics-Accelerators and Beams (PRSTAB), and Reviews of Modern Physics (RMP). The dataset contains 678,916 scientific papers published from 1977 to 2020.

APS has started to use the Physics and Astronomy Classification Scheme (PACS) developed by the American Institute of Physics to classify the fields and sub-fields of research since 1975. There are 441340 articles with the PACS code, published from 1977 to 2015. Most papers published from 1980 to 2010 are labeled with five 6-digital PACS codes. We consider the PACS codes as the categories of scientific research. The PACS codes of a paper demonstrate the recombination of categories. Using the PACS code, we measure knowledge spanning with both word embedding models and network analysis.

**Measures**

We employ word embedding models to embed the categories of research articles into the high-dimensional knowledge space. According to prior research (Levy & Goldberg, 2014), word embeddings are equivalent to pointwise mutual information (PMI). Word embedding models are a particular form of neural language model which aims to embed words into a high-dimensional space (Kozlowski et al., 2019; Mikolov, Chen, et al., 2013; Mikolov, Sutskever, et al., 2013). In this study, each PACS code of APS articles is viewed as a word, and the APS articles are viewed as documents. According to the logic of word embedding models, we distinguish the PACS codes into the center categories and their surrounding categories (Mikolov, Chen, et al., 2013).

1. The input of the neural network model is a sequence of center categories, and the output of the neural network model is the predicted surrounding categories:

2. Each category will be initialized to a random vector with a fixed length of $M$ ($M = 50$) before the training of the neural network models. Thus, we can get an embedding matrix.

3. Using the skip-gram and negative sampling strategies, we train the neural network model, predict the surrounding categories, compare the predicted categories with actual categories to compute the loss, and update the model parameters based on the loss and automatic differentiation.

4. After training the neural network model, we get the updated word embedding matrix.

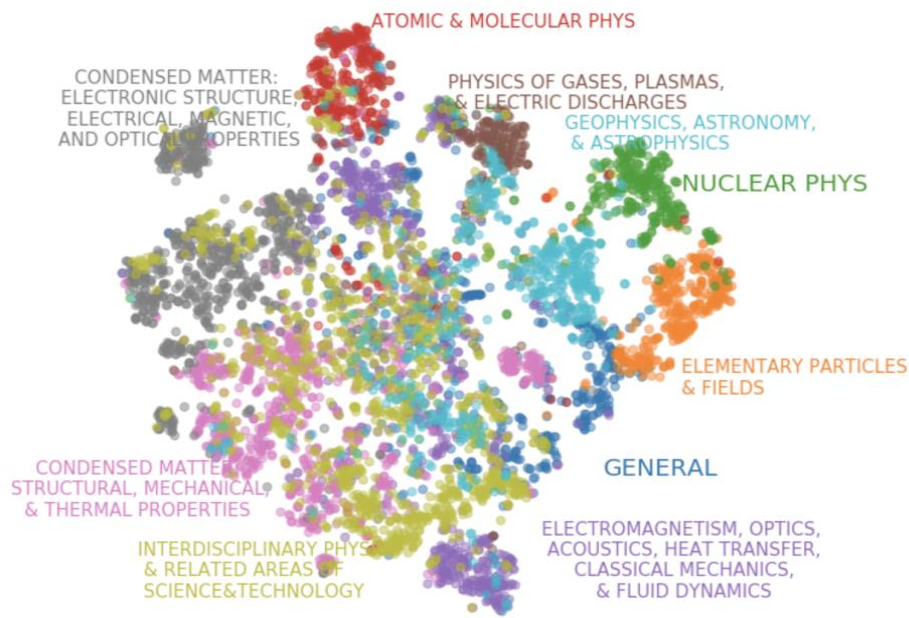

**Figure** 1. The 2-dimensional Knowledge Space of PACS codes

Using word embedding models, we can locate the position of each PACS code within the 50-dimensional knowledge space. To visualize the high-dimensional knowledge space as a two-dimensional picture, we employ the t-distributed stochastic neighbor embedding (t-SNE) algorithm to reduce the dimension of each PACS code's vector from 50 to 2 (Maaten & Hinton, 2008). **Figure** 1 visualizes the 2-dimensional knowledge space of PACS codes after dimension reduction. Each point represents a PACS code, and the color of the points represents which category it belongs to. As shown in **Figure** 1, the distribution of the PACS codes of interdisciplinary physics and related areas is relatively scattered. The scattered distribution of the PACS codes regarding interdisciplinary research is self-evident: interdisciplinary means entering different disciplines. In addition, most PACS codes belonging to one category are perfectly grouped. **Figure** 1 suggests that the knowledge space generated with word embedding models can effectively capture the relative position of different PACS codes.

We further measure internal knowledge spanning (article distance and network distance) and external knowledge spanning (journal distance) based on the results of word embedding models and network analysis (see **Figure** 2). First, as illustrated above, we can use word embedding models to represent the positions of articles, categories, and journals. We define the cosine distance of two given vectors as *1- Cosine Similarity* of these two vectors. Thus, we can measure the scope covered by the categories of an article (article distance) and its distance to the journal in which it is published (journal distance). Second, the PACS codes have a tree-like network structure featured by distinct hierarchies. Thus, we can represent it as a knowledge tree and measure the network distance covered by a given set of PACS codes.

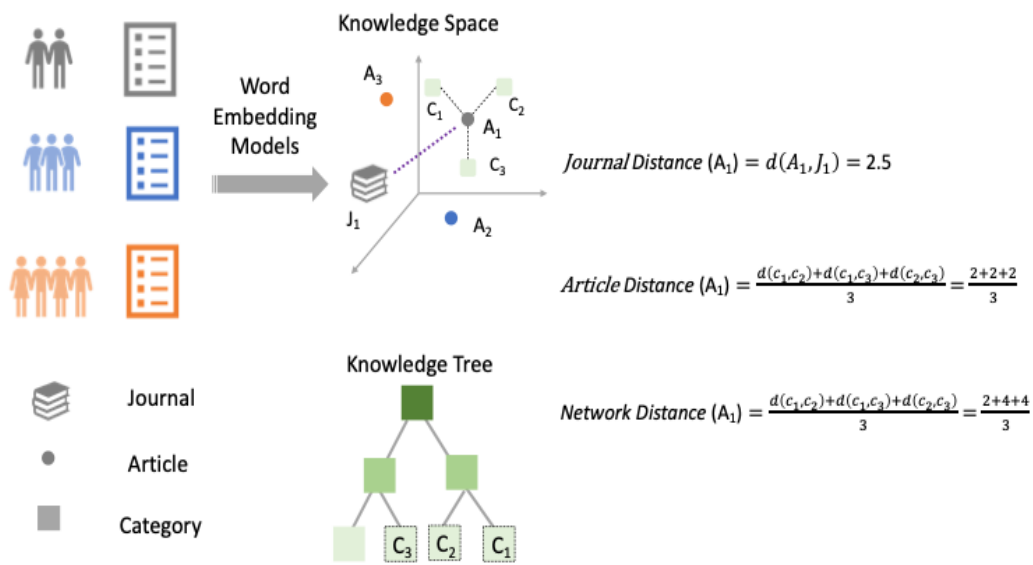

**Figure** 2. Quantifying Knowledge Spanning with Knowledge Space and Knowledge Tree

The tradition is constantly changing and continually reproduced by scientists' strategic choices about what to study and cite (Bourdieu, 1975). Figure 3 visualizes the evolution of APS journals in the knowledge space from 1985 to 2015. Different colors denote different journals. The arrow indicates the chronological order from 1985 to 2015. The broken line shows the positions of the journal over time. To better visualize the direction of the evolution, we smooth the trajectories and plot the smoothed lines. Surprisingly, there seems to be a direction for the evolution of the APS journals. Although the direction of scientific development is brought by the paradigm established by prior scientific revolution or innovation, it is primarily maintained by tradition.

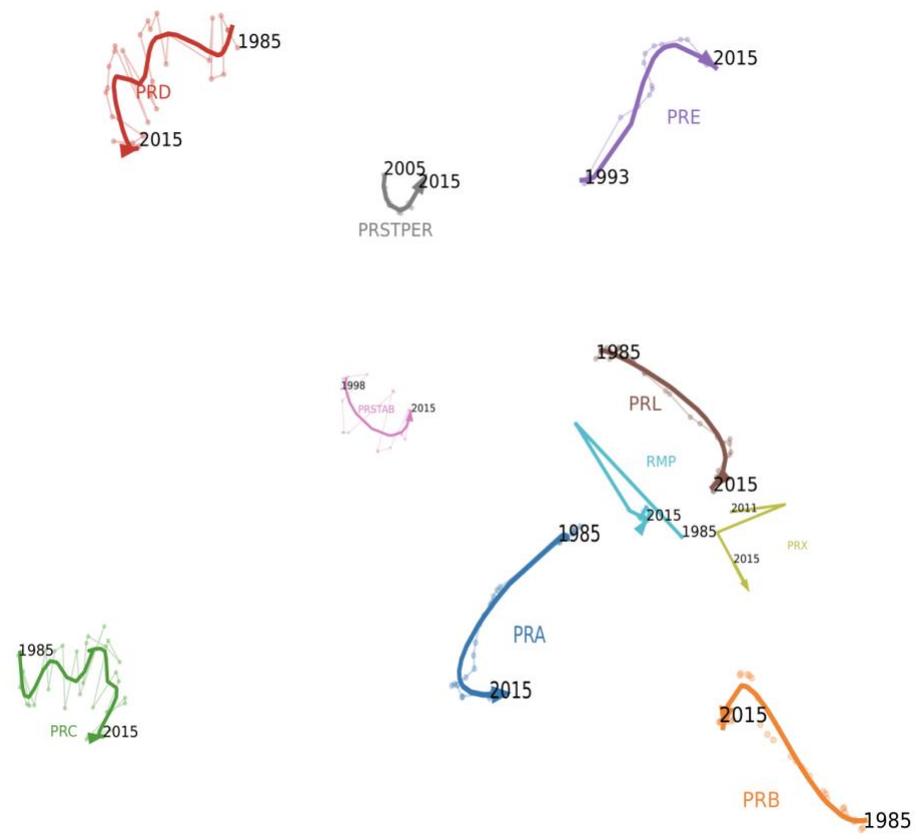

Figure 3. The Evolution of APS Journals in the 2-dimensional Knowledge Space Over Time

**Journal Distance**. In this study, we use the journal distance from paper $i$ to journal $j$ to measure the paper $i$'s external knowledge spanning. Journal distance measures the cosine distance between the focal paper vector and its journal vector in its publication year ($M = 0.24$, $SD = 0.11$). Thus, to calculate journal distance, we need to measure the position of paper $i$ and journal $j$ in year $y$, and y is the year when paper $i$ is published in journal $j$.

First, we can get the category vectors $v_c$ for each PACS code using word embedding models. Suppose that the paper $i$ has $m$ PACS codes. Then, its paper vector $v_i$ can be measured by the mean value of its PACS codes' category vectors $v_{p,i}$, $p \in (1, m)$:

$$v_i = \frac{1}{m} \sum v_{p,i}. \tag{1}$$

Second, for a given journal $j$, suppose that the number of papers published by journal $j$ in the year $y$ is $n$. We identify all these papers and denote their paper vectors as $v_{k,j}$, $k \in (1, n)$. Then, we can calculate the journal vector $v_{j,y}$ of the journal $j$ in the publication year $y$ as the mean value of $v_{k,j}$:

$$v_{j,y} = \frac{1}{n} \sum v_{k,j}. \tag{2}$$

The journal distance from the paper $i$ to journal $j$ can be measured as the cosine distance between $v_i$ and $v_{j,y}$:

$$Journal\ Distance\ (i,j) = 1 - \frac{v_i \cdot v_{j,y}}{|v_i|\ |v_{j,y}|}. \tag{3}$$

**Article Distance**. Article distance measures the breadth covered by the categories of scientific research in the knowledge space (M = 0.23, SD = 0.14). Given paper $i$ has $m$ PACS codes, then there are $\frac{m(m-1)}{2}$ pairs of categories in paper $i$. A pair of category vectors of paper $i$ can be denoted as $v_{p,i}$ and $v_{q,i}$, where $p\ \&\ q \in (1, m)$ and $p \neq q$. Then, the article distance can be calculated as follows:

$$Article\ Distance\ (i) = \frac{2}{m(m-1)} \sum (1 - \frac{v_{p,i} \cdot v_{q,i}}{|v_{p,i}|\ |v_{q,i}|}). \tag{4}$$

**Network Distance**. Network distance measures the length covered by the categories of scientific research on the knowledge tree ($M = 6.96$, $SD = 2.83$). As shown in Figure 2, we can also measure the internal knowledge spanning with network analysis. Given paper $i$ and its $m$ PACS codes, we can denote one of its $\frac{m(m-1)}{2}$ pairs of categories as $p$ and $q$, where $p\ \&\ q \in (1, m)$ and $p \neq q$. Based on the network structure of the knowledge tree, we can calculate the shorted path length $L(p, q)$ and the network distance of paper $i$ can be calculated using the following formula:

$$Network\ Distance\ (i) = \frac{2}{m(m-1)} \sum L(p, q). \tag{5}$$

Note that the PACS code is a 6-digital number, and by adding a root node to represent physics, the tree network of PACS codes represents the hierarchical relationship of six levels. Level 1 represents the root node, that is Physics; Level 2 represents the ten disciplines of Physics (see Figure 1); Level 3 represents the subdisciplines of Physics, and so on. If two categories on level 6 are distributed on two branches growing from the same node on level 3, the network distance between them would be 6. Moreover, if two categories on level 6 are distributed on two branches growing from the same node on level 2, the network distance between them would be 8. The mean value of 6.96 suggests that most categories are located on level 3 or 2. Most scientific research of APS is distributed within a subdiscipline or discipline of Physics. In other words, interdisciplinary spanning is relatively rare.

**Moderating Variables**

    **Team Size**. Team size is the number of scholars who co-published given research ($M = 3.67$, $SD = 2.82$). In our analysis, the largest team size is 25, and the median is 3.

**Dependent Variables**

    **Citation Counts**. Citation counts are the number of papers that cite the focal paper. Since the distribution of citation counts is highly skewed, we transform it with its logarithmic form ($M = 1.99$, $SD = 1.17$)

    **Disruption Percentile**. Following the tradition of prior research (Funk & Owen-Smith, 2017; Wu et al., 2019), we calculate the $D$-score of disruption for each research in the APS dataset:

$$D = \frac{n_i - n_j}{n_i + n_j + n_k}, \tag{6}$$

where $n_i$ is the number of subsequent papers that cites the focal paper, $n_j$ is the number of subsequent papers that cite both the focal paper and its references, and $n_k$ is the number of subsequent papers that only cites the focal paper's references. There exist some variations of the above formula (6). However, according to prior research (Wu et al., 2019), there is no essential difference between them.

Because the distribution of disruption $D$ is also highly skewed, we convert it to its percentile score. Thus, in our analysis, we use the disruption percentile ($M = 48.02$, $SD = 28.15$) to measure the disruptiveness of scientific research. It is also necessary to note that the measure of disruption $D$ tends to be underestimated in the first few years (Lin et al., 2022).

**Findings**

The key puzzlement of this research focuses on the essential tension between tradition and innovation. We report the correlation matrix of the key variables in **Table** 1. First, internal knowledge spanning (article distance and network distance) has a weak negative correlation with external knowledge spanning (journal distance). Thus, covering a large area in the knowledge space does not mean deviating from mainstream research. Second, the two measurements of internal knowledge spanning (article distance and network distance) have a strong correlation ($r(441250) = .82$, $p < .001$), which suggests that the measurement of internal knowledge spanning has good convergent validity. Third, team size positively correlates with internal knowledge spanning and negatively correlates with external knowledge spanning. Fourth, team size has a positive correlation with citation counts and a negative correlation with disruption percentile. Fifth, citation counts have a negative correlation with disruption percentile ($r(441250) = -.32$, $p < .001$).

Table 1. The Correlation Matrix of Key Variables

|  | Journal Distance | Article Distance (log) | Network Distance | Team Size | Citation Counts (log) | Disruption Percentile |
|---|---|---|---|---|---|---|
| Journal Distance | 1.0*** | -0.14*** | -0.12*** | -0.003** | 0.04*** | -0.02*** |
| Article Distance (log) | -0.14*** | 1.0*** | 0.82*** | 0.04*** | 0.01*** | 0.03*** |
| Network Distance | -0.12*** | 0.82*** | 1.0*** | 0.03*** | 0.03*** | 0.0 |
| Team Size | -0.0** | 0.04*** | 0.03*** | 1.0*** | 0.06*** | -0.04*** |
| Citation Counts (log) | 0.04*** | 0.01*** | 0.03*** | 0.06*** | 1.0*** | -0.32*** |
| Disruption Percentile | -0.02*** | 0.03*** | 0.0 | -0.04*** | -0.32*** | 1.0*** |

Note: *p <0.1; **p<0.05; *** p <0.001

We construct multiple linear regression models to formally test the research hypotheses related to citation counts (see **Table** 2). The first hypothesis *H1* asserts that knowledge spanning has an inverted U-shaped impact on citation counts. Three measurements of knowledge spanning have a significant nonlinear influence on citation counts. To further examine the shape of nonlinear influence, we plot the nonlinear influence of knowledge spanning. As shown in **Figure** 4, article distance has an inverted U-shaped influence on citation counts. In comparison, network distance has a positive impact on citation counts, while journal distance has a largely negative impact on citation counts. Thus, *H1* is partially supported.

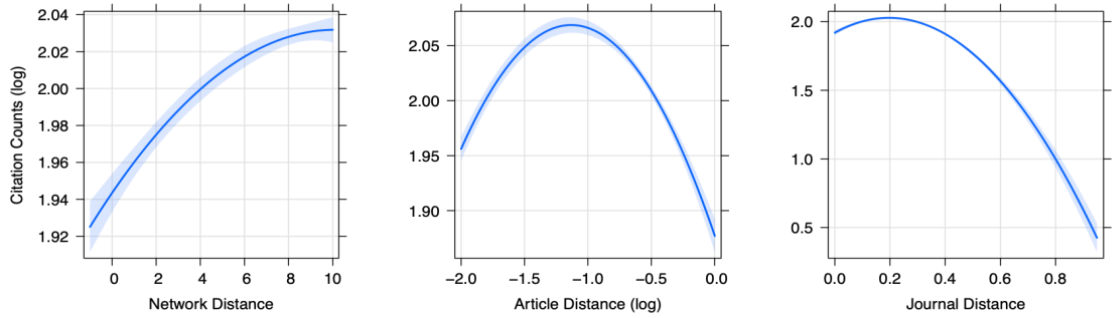

**Figure** 4. The Inverted U-Shaped Influence of Knowledge Spanning on Citation Counts

In addition, *H2a* and *H3a* focus on the impact of team size on citation counts. As **Table 2** shows, team size has a positive main effect on citation counts. Thus, *H2a* is supported. Further, there are significant interactions between team size and knowledge spanning, especially when we include three measurements of knowledge spanning in model 4. We plot the moderation effect to further visually show how team size moderates the effect of knowledge spanning. First, as **Figure** 5 shows, team size can amplify the effect of internal knowledge spanning (network distance and article distance). However, the moderation direction of external knowledge spanning depends on the value of journal distance. When journal distance is smaller than 0.5, team size can amplify the effect of external knowledge spanning; Or else, team size inhibits or reduces the effect of external knowledge spanning. Second, we can also interpret the moderation of team size in terms of the inverted U-shaped curve. Specifically, team size can strengthen the inverted U-shaped relationship between citation counts and network or journal distance. In contrast, team size would flatten the inverted U-shaped relationship between citation counts and article distance. Thus, *H3a* is also supported by the findings.

**Table 2**. Regression Models of Citation Counts

| | Citation Counts (log) | | | |
| --- | --- | --- | --- | --- |
| | Model 1 | Model 2 | Model 3 | Model 4 |
| Number of Pages | 0.039*** | 0.039*** | 0.039*** | 0.039*** |
| Number of Years | 0.012*** | 0.012*** | 0.012*** | 0.012*** |
| Title Length | -0.015*** | -0.015*** | -0.015*** | -0.015*** |
| Network Distance | 35.467*** | | | 71.603*** |
| Network Distance$^2$ | -21.857*** | | | -17.597*** |
| Article Distance (log) | | 16.051*** | | -57.781*** |
| Article Distance (log)$^2$ | | -45.755*** | | -27.084*** |
| Journal Distance | | | -49.678*** | -47.522*** |
| Journal Distance$^2$ | | | -23.880*** | -27.450*** |
| Team Size | 0.026*** | 0.027*** | 0.029*** | 0.029*** |
| Network Distance * Team Size | -3.142*** | | | -10.451*** |
| Network Distance$^2$ * Team Size | -1.140*** | | | 2.870*** |
| Article Distance (log) * Team Size | | 0.064 | | 7.345*** |
| Article Distance (log)$^2$ * Team Size | | 1.461*** | | -2.639*** |
| Journal Distance * Team Size | | | -3.022*** | -3.582*** |
| Journal Distance$^2$ * Team Size | | | -2.857*** | -2.779*** |
| Constant | 2.242*** | 2.244*** | 2.271*** | 2.289*** |
| Adjusted R$^2$ | 0.127 | 0.128 | 0.132 | 0.135 |

Note: * $p<0.1$; ** $p<0.05$; *** $p<0.001$

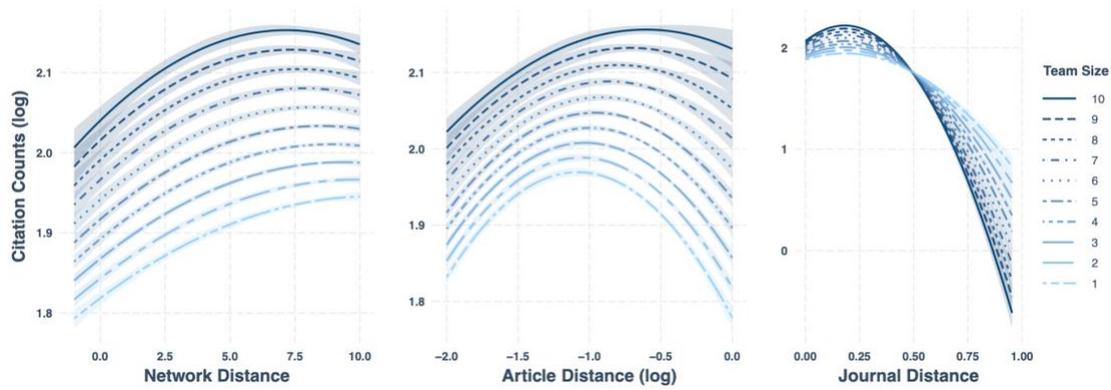

Figure 5. The Moderation Effect of Team Size

To test the research hypotheses related to disruptive innovation, we also construct multiple linear regression models (see **Table** 3). First, according to Model 5-8, the moderation effect of team size is insignificant. Second, team size consistently has a negative impact on the disruption percentile. Therefore, *H2b* is well supported while *H3b* is rejected.

Finally, our operational research question *Q1* concerns the relationship between knowledge spanning and disruptive innovation. As shown in **Table** 3, The square terms of knowledge spanning are generally significant, especially in Model 8. We visually show the relationship between knowledge spanning and disruption percentile in **Figure** 6. Surprisingly, there exists a U-shaped influence of knowledge spanning across three measurements. We have further discussed the implications of these findings in the discussion section.

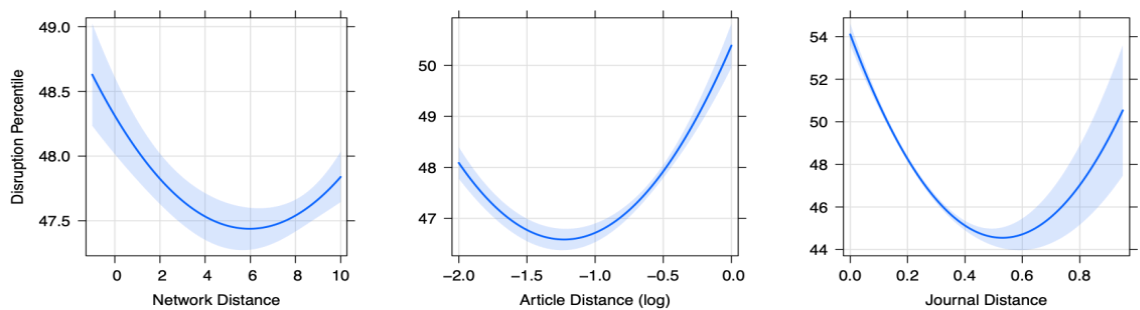

**Figure** 6. The U-Shaped Influence of Knowledge Spanning on Disruptive innovation

Table 3. Regression Models of Disruption Percentile

|  | Disruption Percentile | | | |
| --- | --- | --- | --- | --- |
|  | Model 5 | Model 6 | Model 7 | Model 8 |
| Number of Pages | -0.054*** | -0.054*** | -0.048*** | -0.047*** |
| Years | 0.053*** | 0.052*** | 0.058*** | 0.054*** |
| Title Length | -0.077*** | -0.073*** | -0.079*** | -0.071*** |
| Network Distance | -81.246 |  |  | -457.427*** |
| Network Distance$^2$ | 55.451 |  |  | 159.831* |
| Article Distance (log) |  | 119.694** |  | 363.838*** |
| Article Distance (log)$^2$ |  | 389.974*** |  | 300.389*** |
| Journal Distance |  |  | -906.261*** | -917.052*** |
| Journal Distance$^2$ |  |  | 417.924*** | 430.694*** |
| Team Size | -0.387*** | -0.376*** | -0.409*** | -0.397*** |
| Network Distance * Team Size | 30.751*** |  |  | 9.878 |
| Network Distance$^2$ * Team Size | 10.637 |  |  | 8.933 |
| Article Distance (log) * Team Size |  | 28.230** |  | 28.436 |
| Article Distance (log)$^2$ * Team Size |  | 10.649 |  | 2.942 |
| Journal Distance * Team Size |  |  | 25.863*** | 29.820*** |
| Journal Distance$^2$ * Team Size |  |  | -2.005 | 4.981 |
| Constant | 46.835*** | 46.632*** | 47.362*** | 47.126*** |
| Adjusted R$^2$ | 0.023 | 0.023 | 0.024 | 0.025 |

Note: * p <0.1; ** p<0.05; *** p <0.001

**Discussion and Conclusion**

In summary, this study aims to unpack the essential tension between tradition and innovation proposed by Thomas Kuhn (1977). Our findings suggest that the essential tension holds when we evaluate scientific research by citation counts. Specifically, there is an inverted U-shaped relationship between knowledge spanning and citation counts. Interestingly, network distance has a positive impact on citation counts. In contrast, journal distance has a negative impact on citation counts, especially when the team size is small (see Figure 4 and Figure 5). However, when we evaluate scientific research by disruptive innovation, both lower and higher levels of knowledge spanning can breed disruptive innovation. In contrast, middle-ranged knowledge spanning hinders disruptive innovation. In this sense, the essential tension seems to disappear. Further, our findings on the role of team size reveal that smaller teams have fewer citation counts but more disruptive innovation than larger teams (Wu et al., 2019). In addition, team size moderates the inverted U-shaped relationship between knowledge spanning and citation counts.

Why does network distance fail to capture knowledge spanning's negative impacts on citation counts? On the one hand, interdisciplinary spanning within Physics is relatively rare. As we have analyzed in the method section, the mean value of network distance ($M = 6.96$) suggests that most research is located within a subdiscipline or discipline of Physics. On the other hand, the tree network of categories represents the hierarchical relationship between different categories, but its capacity of representation seems to be relatively limited. In contrast, the article distanced measured with the geometric approach perfectly captures the inverted U-shaped relationship between knowledge spanning and citation counts.

Our findings largely support Kuhn's insight when we evaluate the benefits of knowledge spanning with the currency of scientific credit—citation counts (Kuhn, 1977). According to the idea of the essential tension proposed by Thomas Kuhn (1977), to maximize benefits, scientists need to balance tradition and innovation by choosing the optimal level of knowledge spanning. Guan, Yan, and Zhang (2017) find that knowledge elements' average centrality (i.e., the element's combinatorial opportunities with other elements) in the knowledge network has inverted U-shaped effects on papers' citation counts. Foster and his colleagues (2015) have also examined the essential tension by analyzing the inverted U-shaped relationship between five research strategies (jump, new bridge, repeat bridge, new consolidation, and repeat consolidation) on citation counts. In particular, pure innovation is highly penalized by the research attention in the form of citation counts, while modest fractions of innovation are highly rewarded. In addition, our findings further show that the punishment for external knowledge spanning (i.e., journal distance) is much more severe than internal knowledge spanning (e.g., article distance, see **Figure** 4). In line with Foster et al. (2015), these findings show that the essential tension has a strategic origin in maximizing citation counts, accumulating scientific capital, and occupying a higher position in the scientific field (Bourdieu, 1975). Interestingly, Chen, Arsenault, and Larivière (2015) argue that the top 1% most cited papers exhibit higher levels of interdisciplinarity. Similarly, Chen et al. (2021) find that highly cited papers always exhibit greater variety. One reasonable explanation is that the highly cited papers are featured by a middle-level of knowledge spanning. Compared to highly cited papers, much more scientific research is featured by higher level of knowledge spanning and lower citation counts.

Surprisingly, the essential tension tends to disappear when we use disruptive innovation as an alternative indicator to evaluate science. Kuhn's assertion of the essential tension argues that both tradition and innovation are indispensable (Kuhn, 1977). Confusingly, he opposes tradition and innovation, especially in the phase of normal science. In the seminal work *The Structure of Scientific Revolutions*, Kuhn (1962) largely denies the role of innovation in normal science. Instead, Kuhn (1962) views innovation as the unintended outcome of professionalization. Consequently, scientists are generally disciplined by the professionalization logic in the scientific field organized with citation counts. Against this backdrop, we discover that both lower and higher levels of knowledge spanning can breed disruptive innovation. This discovery reminds us to reconsider the interpretation of the essential tension in terms of disruptive innovation. When evaluating science with disruptive innovation, the relationship between tradition and innovation is not an optimization problem with only one answer. In contrast, middle-ranged knowledge spanning turns out to be harmful to disruptive innovation. Thus, tradition or the dominant paradigm does play a critical role in developing normal science. But Kuhn's narrative about tradition and innovation in normal science could be only partially correct. Our findings highlight the role of the higher-level knowledge spanning in scientific research.

This study has important policy implications. First, using disruptiveness as another currency of scientific credit can better relieve the tension between tradition and innovation. If we evaluate scientific research with citation counts, keeping an optimal distance from the crowd, assembling a larger team, and chasing the hot spots of scientific research would be highly rewarded. On the contrary, if we evaluate the impact with disruptive innovation, scientists can stay either closer to or farther from the crowd in order to achieve more disruptive innovation. Second, we do not advocate using disruptive innovation to replace citation counts. It usually takes a longer time for disruptive research to be recognized and obscure the light of prior research.

We acknowledge the limitations of this study which shed light on future research. First, we quantify knowledge spanning by the PACS code of APS papers. However, APS has stopped using the PACS code, and most other journals generally use the keywords given by the submission system or researchers themselves. Second, similar to the seminal work of Thomas Kuhn (1962, 1977), our study also primarily focuses on the knowledge spanning in the discipline of Physics. In summary, whether this research approach can be generalized to other situations or disciplines deserves more attention in future research.

This study concludes that the essential tension between tradition and innovation would disappear if we change the currency of scientific credit from citation counts to disruptive innovation. From the perspective of disruptive innovation, both lower and higher levels of knowledge spanning can disrupt science. In this sense, there is a binary choice between tradition and innovation. Scientists need to make their choice when they start their scientific research. Although the strategic choice of the optimal level of knowledge spanning can increase the citation counts, it will constrain the capacity of disruptive innovation. In all, disruptive innovation deserves to become another currency of scientific credit.